\documentclass{article}
\usepackage[preprint]{colm2025_conference}
\usepackage{wrapfig}

\usepackage[T1]{fontenc}
\usepackage[utf8]{inputenc}
\usepackage{microtype}
\usepackage{amsmath}
\usepackage{amssymb}
\usepackage{amsfonts}
\usepackage{mathtools}
\usepackage{paracol}
\usepackage{caption}
\usepackage{enumitem}

\usepackage{graphicx}
\usepackage{booktabs}
\usepackage{multirow}
\usepackage{array}
\usepackage{tabularx}
\usepackage{longtable}
\usepackage{makecell}
\usepackage{adjustbox}
\usepackage{colortbl}
\usepackage{float}
\usepackage{subcaption}

\usepackage[table,dvipsnames]{xcolor}
\usepackage[most]{tcolorbox}
\usepackage{xspace}
\usepackage{soul}
\usepackage{ulem}
\usepackage{pifont}
\usepackage{listings}
\usepackage{fancyvrb}
\usepackage{fvextra}
\usepackage{newunicodechar}
\newunicodechar{～}{\textasciitilde}
\usepackage{url}
\usepackage[
  colorlinks=true,
  linkcolor=black,
  citecolor=blue,
  urlcolor=MidnightBlue
]{hyperref}
\usepackage{bookmark}

\definecolor{HardBlue}{RGB}{0,45,120}

\usepackage{tcolorbox}
\usepackage{listings} 
\tcbuselibrary{listings} 
\tcbuselibrary{breakable}

\newtcbinputlisting{\promptbox}[2]{%
  breakable,
  title=#1,
  colback=black!5!white,
  colframe=black!75!black,
  fonttitle=\bfseries,
  listing only,
  listing options={
    basicstyle=\ttfamily\small,
    breaklines=true,
    breakatwhitespace=false,
    breakautoindent=true,
    keepspaces=true
  },
  listing file={#2},
  }
\newtcolorbox{showcase}[2][]{colback=gray!5!white,
  colframe=blue!50!black, 
  title={#2},
  fonttitle=\bfseries,
  sharp corners,
  boxrule=1pt,
  left=2mm, right=2mm, top=1mm, bottom=1mm,
  enhanced,
  breakable,
  #1
}

\renewcommand{\HeaderLeftLogo}{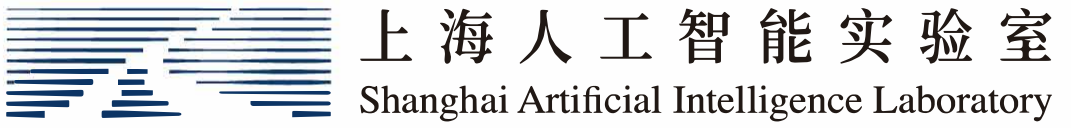}
\renewcommand{\HeaderRightLogo}{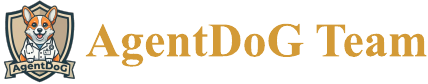}
\newcommand{\frameworkname}{\textsc{EviRepair}}
\newcommand{\coord}{\textsc{Analysis Coordinator}\xspace}
\newcommand{\reach}{\textsc{Reachability Analyzer}\xspace}
\newcommand{\repro}{\textsc{Vulnerability Reproducer}\xspace}
\newcommand{\bound}{\textsc{Boundary Analyzer}\xspace}
\newcommand{\reviewer}{\textsc{Evidence Consistency Reviewer}\xspace}
\newcommand{\rcoord}{\textsc{Repair Coordinator}\xspace}
\newcommand{\synth}{\textsc{Patch Synthesizer}\xspace}
\newcommand{\verifier}{\textsc{Patch Reviewer}\xspace}

\usepackage{geometry}
\title{\centering EviRepair: Evidence-Driven Agents for Reliable Vulnerability Repair}

\author{
    \parbox{0.98\textwidth}{
    \centering
    \normalfont
    \normalsize
    \textbf{Yilian Shi}\textsuperscript{1, 2, *},
    \textbf{Qingyu Liu}\textsuperscript{2,*},
    \textbf{Tianyi Zhou}\textsuperscript{2,*},
    \textbf{Wende Tan}\textsuperscript{3},
    \textbf{Chenyang Si}\textsuperscript{1},
    \textbf{Yueming Lyu}\textsuperscript{1,\textdagger},
    \textbf{Xia Hu}\textsuperscript{2},
    \textbf{Dongrui Liu}\textsuperscript{2,\textdagger}\\[0.35em]
    \textsuperscript{1}School of Intelligence Science and Technology, Nanjing University \\
    \textsuperscript{2}Shanghai Artificial Intelligence Laboratory
    \quad
    \textsuperscript{3}Department of Computing, Imperial College London\\[0.25em]
    \texttt{shiylian@foxmail.com}
    \quad
    \texttt{liuqingyu@pjlab.org.cn}
    \quad
    \texttt{zhoutianyi@pjlab.org.cn}
    }
}

\begin{document}
\maketitle

\begingroup
\renewcommand{\thefootnote}{\fnsymbol{footnote}}
\footnotetext[1]{Equal contribution.}
\footnotetext[2]{Corresponding author.}
\endgroup

\begin{abstract}
Recent advances in large language models (LLMs) have enabled agent-based approaches for repository-level automated vulnerability repair (AVR), showing promising repair capabilities. However, an important issue has received limited attention: \textbf{how can we ensure that the evidence generated during the repair process is reliable enough to support accurate vulnerability understanding and patch generation}? In this paper, we propose \frameworkname{}, an evidence-driven multi-agent framework for repository-level AVR. \frameworkname{} introduces evidence review into the repair process by enabling agents to collaboratively collect vulnerability evidence, verify evidence consistency, and construct a reviewed vulnerability model for patch synthesis and validation. Extensive experiments on PatchEval and SecBench benchmarks with multiple agent frameworks and LLMs demonstrate that \frameworkname{} consistently outperforms existing agent-based baselines, achieving average improvements of 16.37\% and 40.33\% in resolved rates on PatchEval and SecBench, respectively. Codes are available at \url{https://github.com/lly-s/EviRepair}.
\end{abstract}

\section{Introduction}

Automated vulnerability repair (AVR)~\citep{li2025sok} aims to automatically generate patches for software vulnerabilities and has become an important research problem in software security and automated program repair. Early AVR approaches mainly rely on predefined repair templates~\citep{huang2019senx,li2025sok}, program analysis techniques~\citep{gao2021beyond,zhang2022vulnfix}, or search-based strategies~\citep{legoues2012genprog,li2025sok}, which often struggle with real-world vulnerabilities requiring deep understanding of program semantics and repository-level context. As a result, practical vulnerability repair still heavily relies on human expertise, leading to substantial manual effort and limited scalability. Recent advances in large language models (LLMs)~\citep{pearce2023zeroshot,fu2022vulrepair} have opened new possibilities for AVR. By leveraging their code understanding and generation capabilities, LLM-based agents have demonstrated the potential to perform repository-level vulnerability repair with reduced human intervention.

\begin{figure*}[t]
    \centering
    \includegraphics[width=\textwidth]{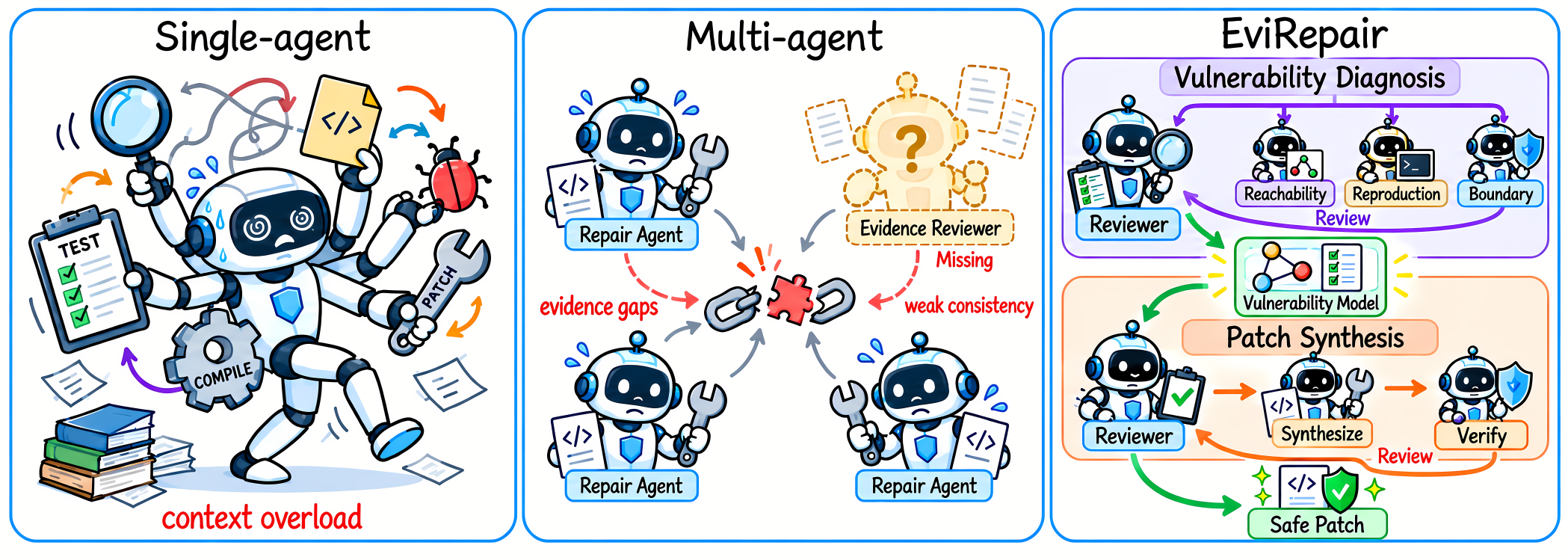}
    \vspace{-10pt}
    \caption{Comparison of single-agent, conventional multi-agent, and EviRepair for automated vulnerability repair. Single-agent methods face context overload by centralizing all tasks, while conventional multi-agent methods omit explicit evidence consistency checks. EviRepair employs multiple agents to alleviate the context burden on individual agents and introduces review agents to ensure consistency.}
    \label{fig:motivation}
\end{figure*}

Despite recent advances in LLM-based AVR systems, generating reliable patches remains challenging. A successful repair requires not only code changes but also evidence that the changes address the underlying vulnerability. However, existing single-agent approaches~\citep{yu2025patchagent,liu2025vuldebugger,kulsum2024vrpilot} perform vulnerability analysis, patch generation, and validation within the same workflow. Consequently, diagnostic evidence is directly reused for repair decisions without a dedicated review stage, allowing incorrect or incomplete vulnerability understanding to propagate through the repair process. Recent multi-agent approaches~\citep{zhang2026sgagent,zhang2026vulnresolver,liu2026memrepair,fu2026agenticrepair} introduce specialized roles to improve repair reliability. However, these frameworks directly aggregate intermediate outputs from different agents. They do not verify evidence consistency before patch generation or use verified evidence for patch validation. As a result, inconsistent or incomplete findings from different agents may be propagated into patch generation. Therefore, existing agent-based AVR frameworks \textbf{generally do not explicitly validate evidence consistency throughout the repair process}.

In this paper, we propose \frameworkname{}, an evidence-driven multi-agent framework for repository-level AVR. \frameworkname{} addresses the lack of evidence validation by introducing evidence review into both vulnerability diagnosis and patch synthesis. It decomposes the repair process into two stages: \textit{Vulnerability Diagnosis} and \textit{Patch Synthesis}, connected through a reviewed \textit{Vulnerability Model}. Given a vulnerability report, the \textit{Analysis Coordinator} in \textit{Vulnerability Diagnosis} assigns complementary analysis tasks to specialized agents to collect vulnerability evidence, including code reachability, reproducibility, security boundaries, and affected behaviors. An Evidence Consistency Review Agent then verifies whether the collected evidence consistently explains the reported vulnerability and determines whether additional analysis is required before constructing the final \textit{Vulnerability Model}. Based on this reviewed model, \textit{Patch Synthesis} plans and generates patches, followed by a patch review against the established security and compatibility criteria.

We conduct extensive experiments on PatchEval~\citep{wei2025patcheval} and SecBench~\citep{lee2025secbench} to evaluate the effectiveness of \frameworkname{}. We compare \frameworkname{} with two mainstream agent frameworks, OpenCode~\citep{opencode2026} and Codex~\citep{openai2026codex} , using two representative LLMs, GLM-5.2~\citep{zai2026glm52} and DeepSeek-4.1-Flash~\citep{deepseek2026v41flash}. Across different agent frameworks and underlying models, \frameworkname{} consistently achieves higher resolved rates than the evaluated baselines, with average improvements of 16.37\% and 40.33\% on PatchEval and SecBench, respectively. We further perform ablation studies to analyze the impact of different design components in \frameworkname{} and investigate their contributions to the overall repair performance.

Our contributions are summarized as follows:
\begin{itemize}[leftmargin=*,itemsep=2pt]
\item We propose \frameworkname{}, an evidence-driven multi-agent framework for repository-level automated vulnerability repair.

\item We design evidence review mechanisms for vulnerability diagnosis and patch synthesis, which validate vulnerability understanding and patch correctness, respectively.

\item We conduct extensive experiments demonstrating that \frameworkname{} consistently outperforms existing agent-based baselines across different frameworks and LLMs.
\end{itemize}

\section{Problem Formulation}

Let $R$ denote a vulnerability report, $P$ denote the target repository, and $E_0$ denote the initially available evidence. In realistic settings, $E_0$ may contain only a natural-language description of the suspected vulnerability. Therefore,  the information required for reliable patching must first be reconstructed from $R$, $P$, and $E_0$.

We represent the repair-critical evidence as
\[
E = \{E_{\mathrm{reach}}, E_{\mathrm{repro}}, E_{\mathrm{oracle}},
E_{\mathrm{boundary}}, E_{\mathrm{preserve}}\},
\]
where the five components describe source-to-sink reachability, executable vulnerability reproduction, runtime observations and expected post-repair behavior, the security boundary, and legitimate behaviors that should be preserved, respectively. These components may be unavailable, incomplete, noisy, or mutually inconsistent. The repair task is to reconstruct a vulnerability model
\[
M = f(R,P,E_0,E)
\]
and synthesize a patch $\Delta$ that blocks the reported unsafe behavior while preserving legitimate program behavior.

\section{\frameworkname{}}

\subsection{Framework overview}

\begin{figure*}[t]
    \centering
    \includegraphics[width=\textwidth]{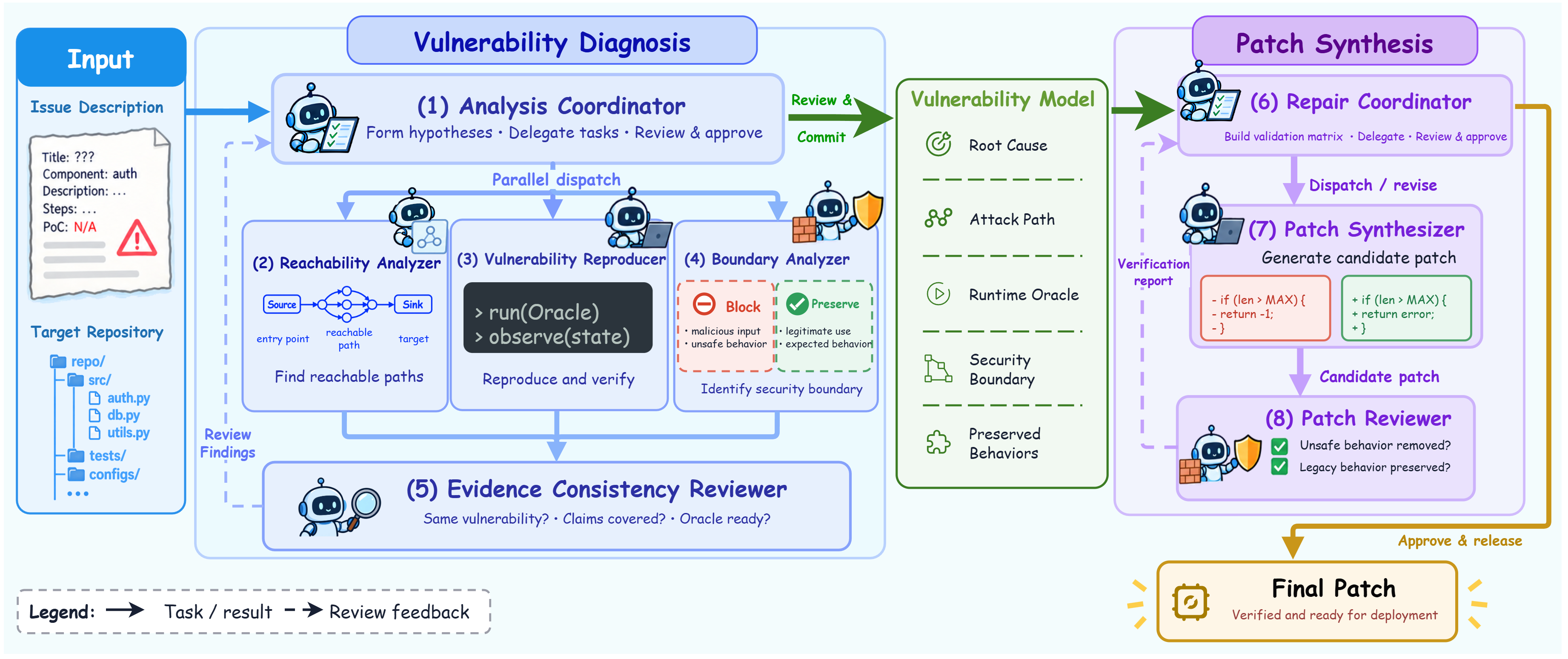}
    \vspace{-10pt}
    \caption{Overview of \frameworkname{}, an evidence-driven multi-agent framework for AVR. Vulnerability Diagnosis gathers complementary vulnerability evidence via specialized agents and validates their consistency to construct a Vulnerability Model. Patch Synthesis treats this model as the specification to generate, validate, and refine candidate patches.}
    \label{fig:framework-overview}
\end{figure*}

As shown in Figure~\ref{fig:framework-overview}, \frameworkname{} is organized as two coordinated multi-agent stages with distinct responsibilities: \textit{Vulnerability Diagnosis} and \textit{Patch Synthesis}. \emph{Vulnerability Diagnosis} (Section~\ref{sec:diagnosis}) collects evidence about the reported vulnerability through independent specialized agents, reviews the evidence for consistency, and assembles it into the Vulnerability Model $\mathcal{M}$ (Section~\ref{sec:vmodel}). \emph{Patch Synthesis} (Section~\ref{sec:synthesis}) treats $\mathcal{M}$ as a fixed specification, producing a patch and verifying it against $\mathcal{M}$. Each agent runs in its own context, and agents exchange results only through structured records with fixed fields.

\subsection{Vulnerability Diagnosis}
\label{sec:diagnosis}
Vulnerability Diagnosis takes a sparse vulnerability report and its target repository as input, and produces the Vulnerability Model $\mathcal{M}$ that guides patch synthesis. Since the report states what is suspected rather than what has been observed, diagnosis treats every claim in it as a hypothesis to be verified, and builds $\mathcal{M}$ only from evidence collected in the repository.

\textbf{Hypothesis formation.} The \coord first reads the report, explores the repository, and forms security hypotheses, each naming a suspected entry point, sink, and vulnerability class. It then dispatches three specialized agents in parallel. Each agent receives the report, the repository, and the hypotheses, and works in its own context without access to the others' findings, so that each investigates one dimension of the vulnerability without interference.

\textbf{Evidence collection.} The three agents collect complementary evidence. The \reach examines whether attacker-controllable input can propagate from the entry point to the sink, and reports the path together with the transformations applied along it. The \repro builds an executable reproducer and a runtime oracle that characterizes the vulnerable behavior before repair and the expected behavior after repair, and runs them on the original code. The \bound derives, from the code around the sink and the vulnerability class, the inputs that must be blocked and the legitimate behaviors that must be retained. Every finding is returned together with the evidence that supports it, such as code locations, analysis results, or execution logs, so that each claim can be checked rather than taken on trust.

\textbf{Review and refinement.} The \reviewer then decides whether these findings are reliable enough to enter $\mathcal{M}$. It checks, partly by re-executing the reproducer and the oracle, that all findings refer to the same vulnerability, that together they cover the claims in the report, and that the oracle is ready for use. Two rules govern this judgment: \emph{agreement among agents is not evidence}, and \emph{a claim supported only by the report is never confirmed}. When a check fails, the \reviewer does not simply reject the findings; it returns a targeted request that names the failed check, the conflicting evidence, and what is missing. The \coord re-dispatches only the responsible agents with this feedback, while retaining the evidence already verified. Diagnosis thus refines its own findings through repeated rounds of review and targeted investigation, until all checks pass or the budget is exhausted. The verified findings are then assembled into $\mathcal{M}$.

\subsection{Vulnerability Model}
\label{sec:vmodel}

Once the Evidence Consistency Reviewer approves, the Analysis Coordinator assembles the validated findings into the Vulnerability Model $\mathcal{M}$, the only diagnostic output passed to Patch Synthesis.  $\mathcal{M}$ consists of five fields:
\begin{itemize}
  \item \textbf{Root Cause:} the location and mechanism of the flaw, with its supporting evidence.
  \item \textbf{Attack Path:} produced by the Reachability Analyzer, the path along which attacker-controllable input travels from the entry point to the sensitive operation, including the transformations applied along the way.
  \item \textbf{Runtime Oracle:} produced by the Vulnerability Reproducer, an executable reproducer together with a check that separately defines what counts as reproduction before patching and as safe behavior after patching.
  \item \textbf{Security Boundary:} produced by the Boundary Analyzer, the violating inputs and their variants that must be blocked after repair.
  \item \textbf{Preserved Behaviors:} produced by the Boundary Analyzer, the legitimate inputs and compatibility requirements that the patch must not break.
\end{itemize}
The last two fields define the acceptance criteria of Patch Synthesis: a patch is accepted only if it removes the unsafe behavior and preserves the legacy behavior. The full schema is given in Appendix~\ref{app:vuln_schema}.

\subsection{Patch Synthesis}
\label{sec:synthesis}

Patch Synthesis turns $\mathcal{M}$ into a verified patch. Every patch is judged against $\mathcal{M}$: it must block the inputs in the Security Boundary and keep the behaviors in the Preserved Behaviors. Since $\mathcal{M}$ has already passed review, this stage does not revise the diagnosis; failures are resolved by revising the patch.

\textbf{Validation matrix.} The \rcoord first translates $\mathcal{M}$ into a validation matrix, the executable form of its acceptance criteria. Each row is a case with an expected outcome: the violating inputs and variants from the Security Boundary must be blocked, the legitimate inputs and compatibility requirements from the Preserved Behaviors must behave as before, and basic checks confirm that the patch applies cleanly and the project still builds. Claims that diagnosis could not confirm are also recorded in the matrix as unverified cases, so that the remaining uncertainty is passed on to repair explicitly instead of being silently ignored. The \rcoord also issues a repair directive that locates the root cause and lists two to four plausible fixes, preferring the one that changes the fewest legitimate behaviors.

\textbf{Patch generation.} Following the directive, the \synth implements a single patch that fixes the root cause rather than filtering the specific reproducer input, so that variants along the same attack path are also blocked. It then executes every row of the validation matrix on the patched repository and records the observed outcomes. Like the diagnostic agents, it therefore returns its result together with the runtime evidence that supports it.

\textbf{Verification and refinement.} The \verifier operates in an isolated context with access to $\mathcal{M}$, the completed validation matrix, and the patch diff, but not to the reasoning of the \synth, so that its judgement remains independent. It approves the patch only if every row matches its expected outcome and the patch is consistent with $\mathcal{M}$: it fixes the flaw at the root cause, removes the unsafe behavior, and preserves the legacy behavior. Passing the matrix alone is not sufficient, since the matrix is built by the system itself; the legitimate inputs and variants guard against patches that merely suppress the observed trigger or reject valid input. When verification fails, the \verifier reports the failing rows and violated conditions, and the \rcoord directs the \synth to revise the patch with this feedback. The cycle repeats until the \verifier approves or the budget is exhausted.

\section{Experiments}
\label{sec:experiments}

Our research questions are as follows:

\begin{description}[leftmargin=1.2cm,style=nextline]
    \item[RQ1] Does \frameworkname{} outperform representative agentic coding systems on vulnerability repair across different models and benchmarks?
    \item[RQ2] To what extent does the performance of \frameworkname{} vary across vulnerability categories?
    \item[RQ3] Is the advantage of \frameworkname{} maintained as repair difficulty increases?
    \item[RQ4] To what extent does each stage of the two-stage multi-agent pipeline contribute to the overall repair performance?
    \item[RQ5] Can structured evidence compensate for the limited reasoning capability of smaller models?
\end{description}

\subsection{Experimental Setup}

\paragraph{Baselines.}
We compare \frameworkname{} against two representative agentic coding systems, Codex v0.153.2~\citep{openai2026codex} and OpenCode v1.17.7~\citep{opencode2026}. All systems use the same underlying LLM, and we vary the backbone among DeepSeek-V4.1-Flash~\citep{deepseek2026v41flash}, GLM-5.2~\citep{zai2026glm52}, and Qwen3.8-27B~\citep{qwen2026qwen38}. All systems run in the same sandboxed environment with identical tool access. For our framework, the two stages have separate time limits: vulnerability diagnosis is limited to 3 hours, and patch synthesis is limited to 2 hours. For OpenCode and Codex, we use a unified total time limit of 5 hours per task.

\paragraph{Benchmarks.} We evaluate on PatchEval and SEC-bench \citep{wei2025patcheval,lee2025secbench}. PatchEval contains 230 CVE repair tasks in Go (83), Python (70), and JavaScript (77), each evaluated dynamically with hidden tests. The agent receives only the vulnerability description and the target repository, without an executable PoC. SEC-bench contains 300 real-world security tasks, including 200 CVE tasks and 100 OSS-Fuzz tasks; its patching tasks additionally provide sanitizer output as auxiliary evidence. The two benchmarks thus cover a description-only setting and an evidence-assisted setting. Network access is disabled during repair, so that no system can retrieve the ground-truth fix from external sources.

\paragraph{Metrics.}
Following the official protocols of both benchmarks, we report the resolved rate (\%Resolved), i.e., the fraction of tasks whose generated patch passes the benchmark's hidden tests. SEC-bench is evaluated under its default medium evaluation setting. Beyond overall numbers, we break down results by programming language on PatchEval (Go / Python / JavaScript) and by task family on SEC-bench (CVE / OSS-Fuzz). For the type- and difficulty-level analyses in RQ2 and RQ3, we average each system's resolved rate over the backbones evaluated for that system to reflect system-level behavior.

\subsection{RQ1: Effectiveness of \frameworkname{}}

\begin{table}[t]
\centering
\caption{Comparison of \%Resolved for vulnerability repair across different agent configurations and backbones on PatchEval and SEC-bench. Numbers in parentheses indicate the total number of vulnerability samples in each subset.
}
\scriptsize
\setlength{\tabcolsep}{4pt}
\begin{tabular}{llcccccccc}
\toprule
\multirow{3}{*}{\textbf{Agent Config.}} & \multirow{3}{*}{\textbf{Backbone}} & \multicolumn{4}{c}{\textbf{PatchEval}} & \multicolumn{3}{c}{\textbf{SEC-bench}} \\
\cmidrule(lr){3-6}
\cmidrule(lr){7-9}
 &
& \makecell{\textbf{Go}\\\textbf{(83)}}
& \makecell{\textbf{Python}\\\textbf{(70)}}
& \makecell{\textbf{JavaScript}\\\textbf{(77)}}
& \makecell{\textbf{Overall}\\\textbf{(230)}}
& \makecell{\textbf{CVE}\\\textbf{(200)}}
& \makecell{\textbf{OSS-Fuzz}\\\textbf{(100)}}
& \makecell{\textbf{Overall}\\\textbf{(300)}} \\
\midrule
\multirow{3}{*}{Codex~\citep{openai2026codex}}
& DeepSeek-4.1-Flash
& 74.70\% & 81.43\% & 79.22\% & 78.26\% 
& 64.50\% & 90.00\% & 73.00\% \\

& GLM-5.2
& 65.06\% & 64.29\% & 72.73\% & 67.39\% 
& 49.00\% & 88.00\% & 62.00\% \\

& Qwen3.8-27B
& 55.42\% & 71.43\% & 67.53\% & 64.35\% 
& 41.50\% & 75.00\% & 52.67\% \\

\midrule

\multirow{3}{*}{OpenCode~\citep{opencode2026}}
& DeepSeek-4.1-Flash
& 85.54\% & 91.43\% & \textbf{87.01\%} & 87.83\% 
& 88.00\% & 94.00\% & 90.00\% \\

& GLM-5.2
& 49.40\% & 64.29\% & 70.13\% & 60.87\% 
& 71.00\% & 90.00\% & 77.33\% \\

& Qwen3.8-27B
& 66.27\% & 65.71\% & 68.83\% & 66.96\% 
& 77.50\% & 93.00\% & 82.67\% \\

\midrule

\multirow{3}{*}{\frameworkname{}}
& DeepSeek-4.1-Flash
& \textbf{91.57\%} & \textbf{94.29\%} & 81.82\% & \textbf{89.13\% }
& \textbf{95.50\%} & \textbf{98.00\%} & \textbf{96.33\%} \\

& GLM-5.2
& 69.88\% & 77.14\% & 72.73\% & 73.04\%
& 93.00\% & 96.00\% & 94.00\% \\

& Qwen3.8-27B
& 75.90\% & 78.57\% & 71.43\% & 75.22\% 
& 90.00\% & 87.00\% & 89.00\% \\

\bottomrule
\end{tabular}
\label{tab:main-results}
\end{table}


\begin{figure*}[t]
    \centering
    \includegraphics[width=\textwidth]{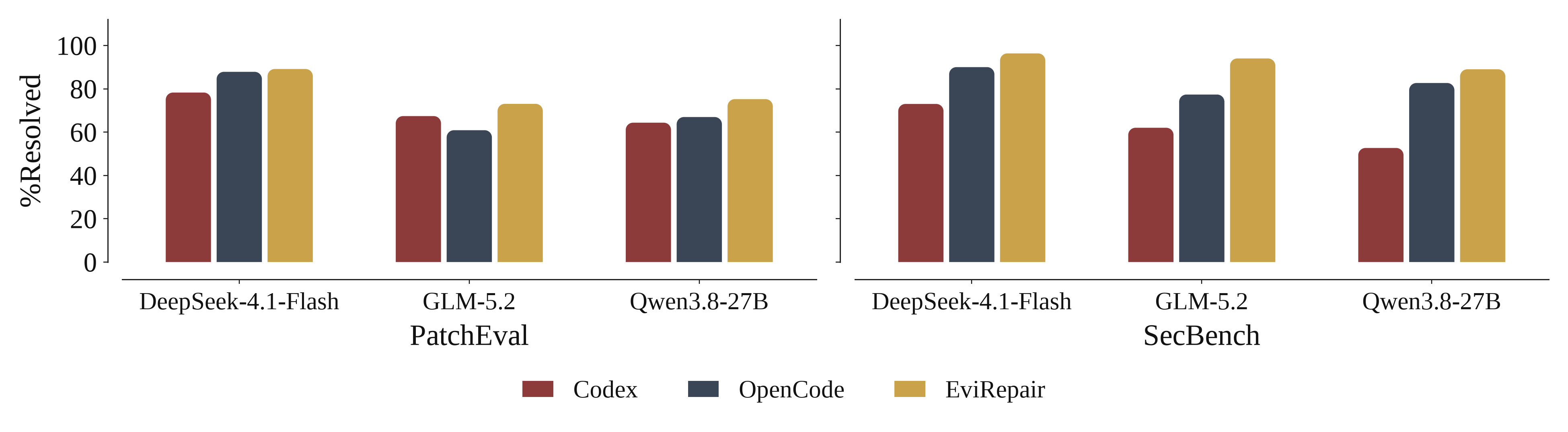}
    \vspace{-10pt}
    \caption{Resolved rate on PatchEval and SEC-bench across three backbones. The numbers above the bars indicate the absolute improvement of \frameworkname{} over each baseline.}
    \label{fig:bar-result}
\end{figure*}

Table~\ref{tab:main-results} shows that \frameworkname{} achieves the highest overall resolved rate on both benchmarks under all three backbones. Averaged over the backbones, it resolves 79.13\% of PatchEval tasks, compared with 70.00\% for Codex and 71.89\% for OpenCode, and 93.11\% of SEC-bench tasks, compared with 62.56\% and 83.33\%. Relative to the stronger baseline, OpenCode, this reduces the fraction of unresolved tasks by about 26\% on PatchEval and 59\% on SEC-bench.

The advantage holds consistently across backbones and subsets. On PatchEval, \frameworkname{} exceeds the strongest baseline by 1.30, 5.65, and 8.26 percentage points with DeepSeek-4.1-Flash, GLM-5.2, and Qwen3.8-27B; on SEC-bench, the margins are 6.33, 16.67, and 6.33 points. Across the 21 comparisons between systems on the same backbone and subset, \frameworkname{} achieves the best result in 18 and ties in one; the only exceptions are JavaScript with DeepSeek-4.1-Flash and OSS-Fuzz with Qwen3.8-27B, where OpenCode performs better.

The largest gains appear on tasks that require understanding the vulnerability rather than locating a crash. On the CVE subset of SEC-bench, \frameworkname{} averages 92.83\%, outperforming Codex by 41.17 points and OpenCode by 14.00 points, whereas on the OSS-Fuzz subset all systems already perform strongly and the margins are smaller. OSS-Fuzz tasks come with sanitizer reports that pinpoint the crashing location, so little additional evidence is needed. CVE tasks instead require establishing the attack path and the security boundary from the report, which is precisely the evidence that Vulnerability Diagnosis reconstructs.

\subsection{RQ2: Performance across vulnerability types}
\begin{figure*}[t]
    \centering
    \includegraphics[width=\textwidth]{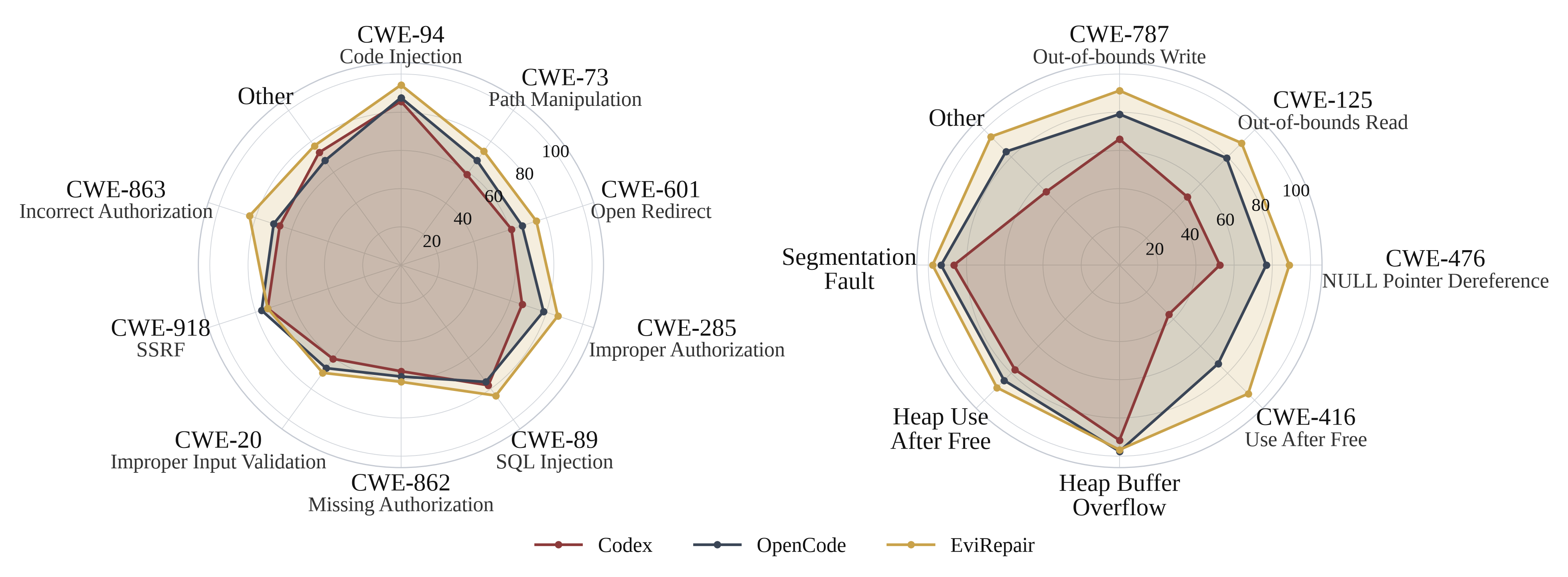}
    \vspace{-18pt}
    \caption{Resolved rate across vulnerability categories on PatchEval and SEC-bench. Each system is averaged over the three backbones.}
    \label{fig:radar_vuln_type}
    \vspace{-10pt}
\end{figure*}

We group tasks by vulnerability category (Figure~\ref{fig:radar_vuln_type}). Categories with more than 10 tasks are shown individually, and the remaining categories are merged into \emph{Other}. For PatchEval, we use the primary CWE in the official metadata; for SEC-bench, we use the primary weakness assigned by NVD for CVE tasks and the root-cause category in the official OSS-Fuzz crash reports.
 
On PatchEval, the largest gains appear on authorization flaws, namely incorrect authorization (CWE-863) and improper authorization (CWE-285), followed by path manipulation (CWE-73) and open redirect (CWE-601). Repairing these flaws requires deciding precisely which requests should be permitted and which rejected, which is the security boundary that the \bound establishes before any patch is written. In contrast, all systems perform similarly on code injection (CWE-94) and SQL injection (CWE-89), whose fixes follow well-known patterns, and \frameworkname{} shows no advantage on server-side request forgery (CWE-918). A likely reason is that reproducing SSRF requires outbound requests, which cannot be observed in our network-isolated sandbox, so the runtime oracle for this category is weaker.
 
On SEC-bench, \frameworkname{} leads in every category. The gains are largest on use-after-free (CWE-416) and NULL pointer dereference (CWE-476), and smallest on heap buffer overflow, where all systems perform strongly. For temporal and null-state errors, the crash reported by the sanitizer typically occurs far from the root cause, such as the point where an object is freed or a pointer is left unset, and locating that point requires tracing the attack path. For buffer overflows, the crash site usually coincides with the location to fix, so the sanitizer report alone already provides most of the necessary evidence.
 
Across both benchmarks, the pattern is consistent: \frameworkname{} helps most when the location of the fix differs from where the vulnerability manifests, or when the fix depends on an explicit boundary between permitted and forbidden behavior.

\subsection{RQ3: Performance across repair difficulty}

\begin{figure*}[t]
    \centering
    \includegraphics[width=\textwidth]{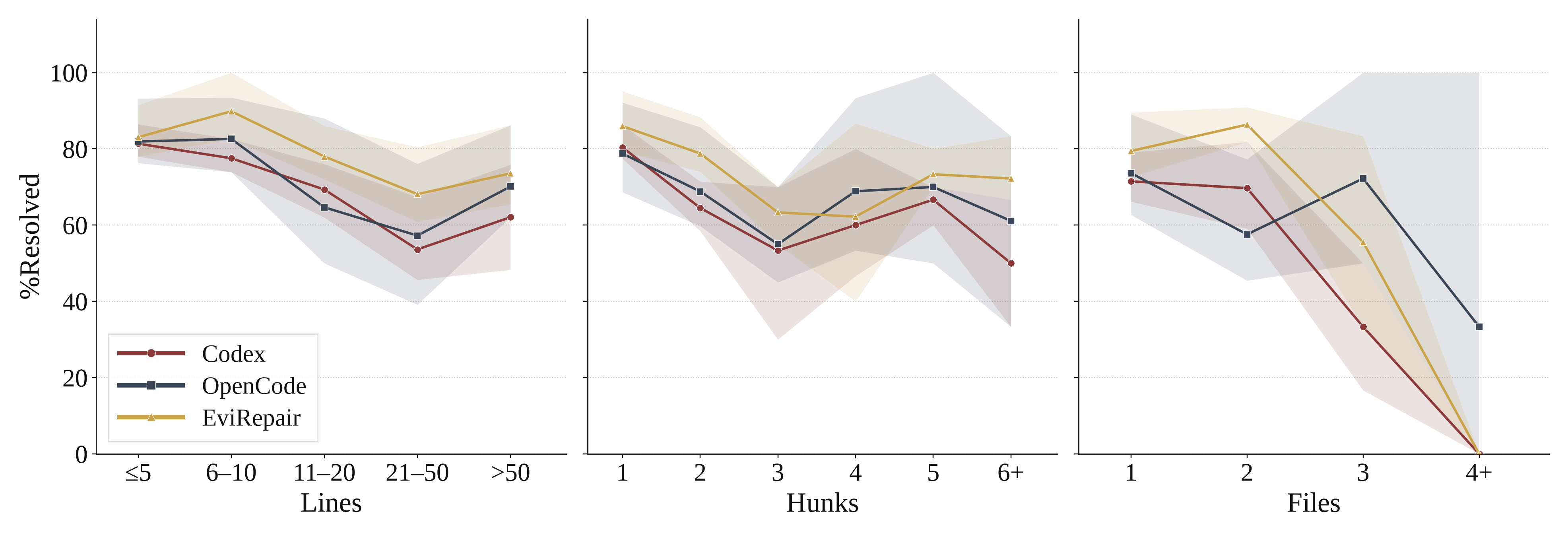}
    \vspace{-18pt}
    \caption{Resolved rate across repair difficulty dimensions on PatchEval: number of changed lines, hunks, and files per ground-truth patch. Lines show the average over three backbones; shaded regions indicate the min--max range.}
    \label{fig:line_difficulty}
\end{figure*}

The advantage of \frameworkname{} is largely maintained as repair difficulty increases, with the exception of the few repairs that span many files (Figure~\ref{fig:line_difficulty}). All systems degrade as ground-truth patches grow larger, but \frameworkname{} degrades more slowly. On the lines dimension, it leads or ties in every bin, and the margin is widest on larger patches: in the 21--50 bin, it reaches roughly 72\%, compared with about 56\% for the baselines. On the hunks dimension, it leads in all bins except the 4-hunk bin, and its lead is largest on the hardest bin of six or more hunks. On the files dimension, it clearly outperforms both baselines on single- and two-file patches. Larger patches typically involve longer attack paths and more intermediate transformations, where an explicitly reconstructed attack path and security boundary are most useful for deciding where and how to fix.

The only setting in which the advantage does not hold is repairs spanning three or more files, which account for only 7 of the 230 tasks. On these tasks, OpenCode performs best, and \frameworkname{} resolves none of those requiring four or more files. Given the small sample, these estimates are unstable, but the pattern is consistent with the structure of the Vulnerability Model, which centers on a single root cause and attack path; fixes that must coordinate changes across several files fall partly outside this representation. Extending the model to multiple interacting root causes is a natural direction for future work.

\subsection{RQ4: Ablation Study}

\begin{table}[t]
\centering
\small
\caption{Ablation study on the two-stage multi-agent pipeline. All experiments use GLM-5.2 as the backbone and are evaluated on PatchEval. Single-Agent Diagnosis/Synthesis uses a single agent in the vulnerability diagnosis/patch synthesis stage while keeping multi-agent for the other stage. Single-Agent E2E uses one agent to perform the entire vulnerability diagnosis and patch synthesis process.}
\label{tab:ablation-results}
\begin{tabular}{lcccc}
\toprule
\textbf{Configuration} 
& \makecell{\textbf{Overall}\\\textbf{(230)}}
& \makecell{\textbf{Go}\\\textbf{(83)}}
& \makecell{\textbf{Python}\\\textbf{(70)}}
& \makecell{\textbf{JavaScript}\\\textbf{(77)}} \\
\midrule
Single-Agent Diagnosis & 69.13\% & 67.47\% & 67.14\% & \textbf{72.73\%} \\
Single-Agent Synthesis & 67.83\% & \textbf{69.88\%} & 65.71\% & 67.53\% \\
Single-Agent E2E & 60.87\% & 49.40\% & 64.29\% & 70.13\% \\
\midrule
\textbf{\frameworkname{}} & \textbf{73.04\%} & \textbf{69.88\%} & \textbf{77.14\%} & \textbf{72.73\%} \\
\bottomrule
\end{tabular}
\end{table}

Both stages are necessary: replacing either stage with a single agent degrades performance, and replacing both causes the largest drop (Table~\ref{tab:ablation-results}). Replacing the multi-agent diagnosis with a single agent reduces the resolved rate from 73.04\% to 69.13\% ($-$3.91 points), replacing the multi-agent synthesis reduces it to 67.83\% ($-$5.21 points), and replacing both reduces it to 60.87\% ($-$12.17 points).

The loss from replacing both stages exceeds the sum of the losses from replacing each stage alone (9.12 points). Removing either stage costs relatively little as long as the other remains multi-agent, whereas removing both causes a much larger drop. This indicates that the two stages reinforce each other: both introduce an independent check before a result is accepted, of the evidence in diagnosis and of the patch in synthesis, so that an error missed by one stage can still be caught by the other. A representative evidence-driven trajectory illustrating this two-stage interaction is provided in Appendix~\ref{app:case_study}.

\subsection{RQ5: Effect of Backbone Capability}

Structured evidence largely compensates for a smaller backbone: with the smallest backbone, \frameworkname{} outperforms baselines running on a larger one. Qwen3.8-27B is the smallest backbone in our study and yields the weakest baseline performance on both benchmarks. With this backbone, \frameworkname{} resolves 75.22\% of PatchEval tasks, exceeding both baselines not only on Qwen3.8-27B but also on the larger GLM-5.2 (67.39\% and 60.87\%). On SEC-bench, it reaches 89.00\%, outperforming every baseline configuration on GLM-5.2 and nearly matching the strongest baseline overall, OpenCode with DeepSeek-4.1-Flash (90.00\%).

The benefit of structured evidence grows as the backbone weakens. On PatchEval, the margin of \frameworkname{} over the strongest baseline on the same backbone increases from 1.30 points with DeepSeek-4.1-Flash to 5.65 with GLM-5.2 and 8.26 with Qwen3.8-27B: the less capable the backbone, the more the evidence reconstructed by Vulnerability Diagnosis determines whether the repair succeeds. This trend is less pronounced on SEC-bench, where sanitizer output already supplies part of this evidence to all systems.

\section{Related Work}

\paragraph{Repair Paradigms.}
Automated vulnerability repair is split into traditional and agent-based paradigms.Traditional methods use predefined correctness signals, generating and filtering patches with handwritten rules, tests and historical fixes \citep{legoues2012genprog,huang2019senx,gao2021beyond,chen2023vrepair,fu2022vulrepair,zhou2024vulmaster}, yet their limited modeling capacity hinders the handling of program dependencies and runtime behaviors in large repositories. LLM-based single-agent repair mitigates this by combining code retrieval, tool use and execution feedback for iterative localization and patch validation \citep{bouzenia2025repairagent,yu2025patchagent}. Still, consolidating diverse reasoning in one context creates tangled information and bias, hurting patch quality. Multi-agent repair addresses this through task specialization. Some works deploy expert agents to interpret vulnerabilities from code, traces and repository history, including AutoPatch and AgenticRepair \citep{seo2025autopatch,fu2026agenticrepair}. Others derive security constraints and repair requirements via multi-agent reasoning, such as VulnResolver and ContraFix \citep{zhang2026vulnresolver,liu2026contrafix}. These studies confirm that multi-agent architectures with decoupled contexts and joint evidence analysis are promising for complex vulnerability repair.

\paragraph{Vulnerability Repair Datasets.}
Datasets differ in validation design and supplied evidence. Static code-pair datasets such as Big-Vul, CVEfixes and MegaVul evaluate patches against vulnerable–fixed code pairs \citep{fan2020bigvul,bhandari2021cvefixes,ni2024megavul}, but cannot confirm whether a patch removes exploitable behavior while preserving normal functionality. Executable datasets remedy this by combining real project revisions and security evidence. Vul4J, Vul4C and VulnRepairEval provide executable evaluation workflows for different languages and strengthen validation with containerized differential execution \citep{bui2022vul4j,hu2025sok,wang2025vulnrepaireval}. Recent progressive repair datasets reduce accessible vulnerability information to mimic real, challenging conditions. PatchEval, SE-bench, CyberGym-E2E and CVE-Bench impose different information constraints, ranging from text-only vulnerability descriptions to black-box settings with minimal code and build materials \citep{wei2025patcheval,lee2025secbench,shi2026cybergyme2e,wang2025cvebench}. Collectively, they show that the availability of vulnerability evidence strongly controls task difficulty and repair performance.

Recent progressive repair datasets identify insufficient vulnerability evidence as a key practical bottleneck, while single-agent and multi-agent paradigms each have complementary strengths and limitations. Building on these findings, this paper investigates evidence reconstruction from sparse vulnerability reports, drawing on multi-agent specialization for focused subtask reasoning while using cross-evidence validation to realize the global consistency targeted in single-agent designs.

\section{Limitations}

The evaluation is limited to PatchEval and SEC-bench, which may not reflect the full diversity of vulnerabilities, languages, repositories, and deployment environments. In addition, security assessment is restricted to benchmark-level validation, so the reported results do not guarantee complete vulnerability elimination, absence of regressions, or long-term maintainability. The multi-agent design also incurs extra model calls, latency, and monetary costs. Future work should evaluate \frameworkname{} in broader real-world settings, strengthen security and regression assessment, and further investigate the trade-off between repair quality and resource cost.

\section{Conclusion}

In this paper, we present \frameworkname{}, an evidence-driven multi-agent framework for repository-level automated vulnerability repair from sparse reports. \frameworkname{} separates Vulnerability Diagnosis from Patch Synthesis and connects them through a reviewed Vulnerability Model. Experiments on PatchEval and SEC-bench show that \frameworkname{} consistently outperforms representative baselines, with average resolved-rate improvements of 16.37\% and 40.33\%, respectively. Ablation studies further demonstrate the value of stage decomposition and independent evidence review, highlighting the importance of evidence-grounded coordination for reliable vulnerability repair. We hope that \frameworkname{} and our findings can facilitate future research on reliable, evidence-grounded automated vulnerability repair.

\section*{AI use statement}

Large language models were engaged exclusively for text refinement, including writing clarity and language editing. Theoretical analysis, research methodology, experimental design, and result interpretation were all independently completed by the authors without reliance on such tools.

\section*{Ethics statement}

This work studies automated vulnerability repair using publicly available vulnerability-repair benchmarks and controlled software repositories. The experiments do not involve human subjects, private or personally identifiable data, or unauthorized testing of live systems. All program analysis, vulnerability reproduction, and patch validation were conducted within the designated benchmark or experimental environments.

Automated vulnerability repair has potential dual-use implications because techniques for reproducing and analyzing vulnerabilities may also be misused against vulnerable software. To reduce these risks, our evaluation focuses on publicly available benchmark tasks and defensive repair objectives. We do not claim that benchmark success guarantees the elimination of all exploit paths in real-world software, and generated patches should be reviewed and validated before deployment.

The authors followed applicable software, dataset, and benchmark licenses and take responsibility for the responsible use and reporting of the methods and results presented in this paper.

\section*{Reproducibility Statement}
The code and selected experimental resources for \frameworkname{} are publicly available at \url{https://github.com/lly-s/EviRepair}. The repository includes the implementation of the \frameworkname{} framework, agent prompts, structured communication schemas, and scripts for running the automated vulnerability repair and benchmark evaluation pipelines on PatchEval and SEC-bench. The accompanying README provides instructions for preparing the execution environment, configuring model endpoints, obtaining and specifying the required benchmark data and repository snapshots, and running the repair and evaluation workflows. Additional experimental details, including the configurations used for evaluation, are documented in the paper.

\bibliographystyle{colm2025_conference}
\bibliography{latex/refs}

\appendix
\section{Full Schema of Vulnerability Model}
\label{app:vuln_schema}
\promptbox{JSON Schema for Vulnerability Model}{template/vulnerability_schem.json}

\section{A Case Study of \frameworkname{}}
\label{app:case_study}

\begin{figure}[ht]
    \centering
    \includegraphics[width=\textwidth]{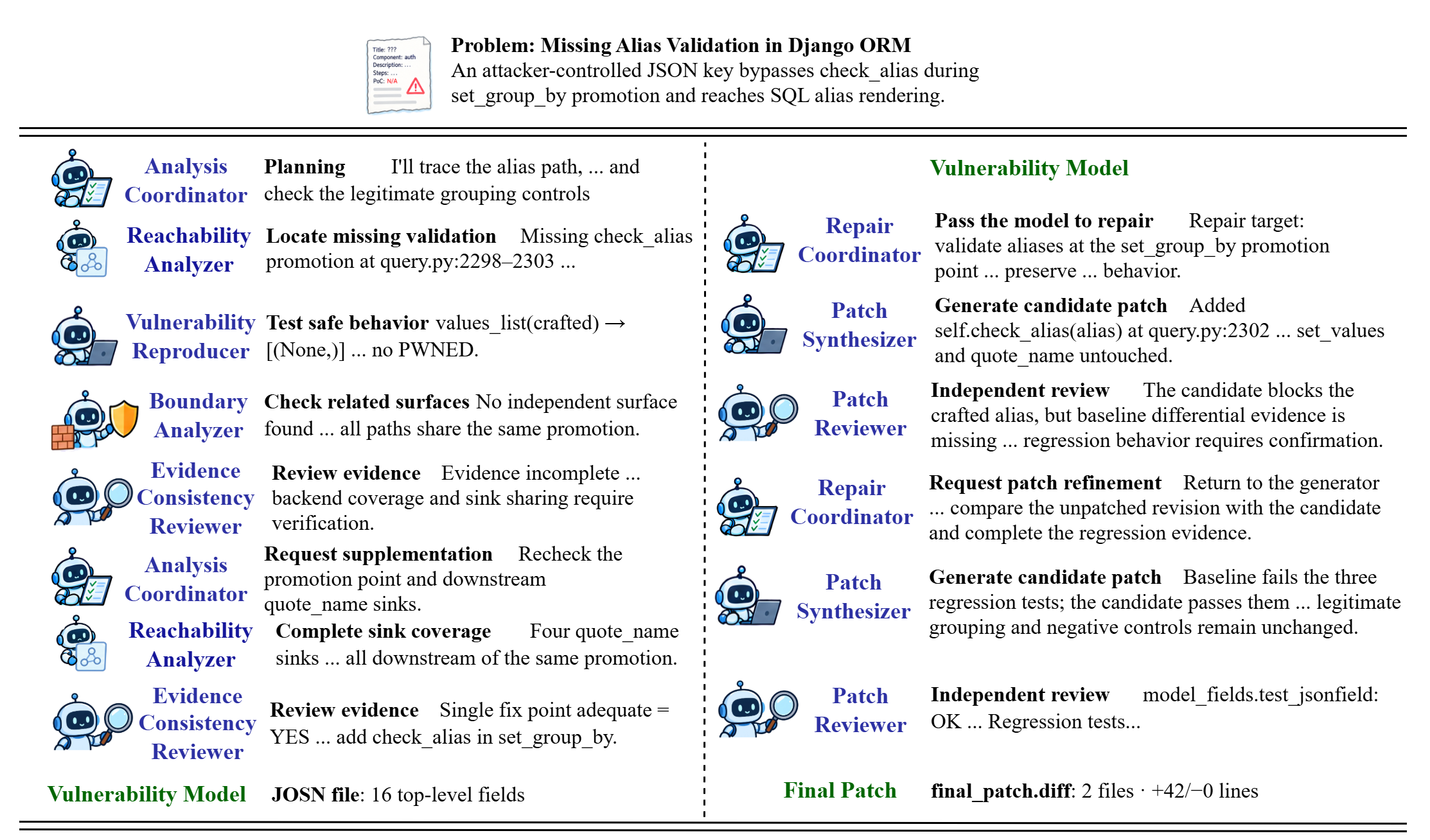}
    \caption{A \frameworkname{} Case Study of Evidence-Driven Two-Stage Vulnerability Repair.}
    \label{fig:case_study}
\end{figure}

\section{Prompt}
Due to space limitations, we only present the reviewer prompts for the two stages here. The remaining prompts can be found in our repository.
\promptbox{Evidence Consistency Reviewer System Template}{template/analysis_mas_workers_evidence_review_worker.system.txt}
\promptbox{Evidence Consistency Reviewer User Template}{template/analysis_mas_tasks_evidence_review_worker.task.txt}
\promptbox{Patch Reviewer System Template}{template/repair_mas_patch_review.system.txt}
\promptbox{Patch Reviewer User Template}{template/repair_mas_tasks_patch_reviewer.task.txt}

\end{document}